\documentclass[cmp]{svjour}
\usepackage{amssymb}
\usepackage{amsfonts}

\def\be{\begin{eqnarray}}
\def\ee{\end{eqnarray}}
\def\bee{\begin{eqnarray*}}
\def\eee{\end{eqnarray*}}
\newtheorem{thm}{Theorem}

\newtheorem{lem}{Lemma}

\newtheorem{defn}{Definition}

          \def\tr{\hbox{Tr}}

\def\bra{\langle}
\def\ket{\rangle}

\def\ot{\otimes}

\begin{document}
\title{Multiplicativity of Accessible Fidelity and
Quantumness for Sets of Quantum States}
\titlerunning{Multiplicativity of accessible fidelity and quantumness}
\author{K.~M.~R.~Audenaert\inst{1}\and C.~A.~Fuchs\inst{2,3}\and 
C.~King\inst{3,4}\and A.~Winter\inst{5}}
\institute{
University of Wales, Bangor \\
School of Informatics \\
Bangor (Gwynedd) LL57 1UT, Wales \\
\email{kauden@informatics.bangor.ac.uk}
\and
Bell Labs, Lucent Technologies \\
600-700 Mountain Ave.\\
Murray Hill, NJ 07974, USA\\
\email{cafuchs@research.bell-labs.com} \and
Communication Networks Research Institute \\
Dublin Institute of Technology \\
Rathmines Road \\
Dublin 6, Ireland
\and
Department of Mathematics \\
Northeastern University\\
Boston, MA 02115, USA\\
\email{king@neu.edu}
\and
Department of Computer Science\\
University of Bristol\\
Merchant Venturers Building\\
Woodland Road\\
Bristol BS8 1UB, England\\
\email{winter@cs.bris.ac.uk}
}
\date{1 August 2003}
\def\makeheadbox{}
\maketitle
\begin{abstract}
Two measures of sensitivity to eavesdropping for alphabets of quantum
states were recently introduced by Fuchs and Sasaki in {\tt
quant-ph/0302092}.  These are the accessible fidelity and
quantumness. In this paper we prove an important property of both
measures: They are multiplicative under tensor products.  The proof
in the case of accessible fidelity shows a connection between the
measure and characteristics of entanglement-breaking quantum
channels.
\end{abstract}
\section{Introduction and statement of results}
The security of quantum cryptography relies on the notion that any
measurement on a quantum system causes a disturbance to it, thereby
revealing the presence of an eavesdropper. However the idea that
`measurement causes disturbance' must be applied carefully in order
to be useful. For example, given a state $| \psi \ket$, the
measurement which projects onto $| \psi \ket$ and its orthogonal
complement causes no disturbance to the state. Furthermore, if a
signal is encoded using orthogonal states for different letters in an
alphabet, then an eavesdropper can gain complete information by
projecting onto those states, again without causing any changes in
the signal. So in order to be successfully exploited for quantum
cryptography (for example as in \cite{BB84}), an encoding scheme must
use an ensemble of nonorthogonal signal states to prevent a
disturbance-free measurement. In other words, the sender cannot use a
classical ensemble of states to implement quantum cryptography.

Thus for purposes of implementing quantum cryptography some ensembles
are better than others.  This raises the question of trying to
quantify the `amount of quantumness' in an ensemble of states. We
will address one aspect of this question using the approach
introduced in the paper~\cite{fuchs}. (For a different approach, see
\cite{Hayden02}.)  The idea of the present approach is to consider
the transmission of an ensemble of states from a sender to a
receiver, and to see how easily an eavesdropper can be detected at
participating in an intercept/resend strategy. Specifically, suppose
that the sender draws states randomly from an ensemble ${\cal E} =
\{p_i, |\psi_i \ket \}$. After transmission the receiver obtains the
ensemble ${\cal E}' = \{p_i, | \psi_i \ket' \}$. In the absence of
noise or an eavesdropper, these ensembles should have fidelity equal
to 1.  Recall that the fidelity is given by
\be\label{def:F}
F = \sum_{i} p_i \,\, \big| \bra \psi_i | \psi_i' \ket \big|^2\;.
\ee

Now suppose that the eavesdropper is allowed to make any measurement
on the intercepted states, that is any fixed POVM $\{E_b \}$ can be
applied. Based on the result of this measurement, the eavesdropper
substitutes any other state $| \phi_b \ket$ in place of $| \psi_i
\ket $ and sends this on to the receiver. The fidelity between the
original ensemble and this new ensemble is
\be\label{def:Fnew}
F' = \sum_{i} \sum_{b} p_i \,\, \bra \psi_i| E_b | \psi_i \ket \, \,
\big| \bra \psi_i | \phi_b \ket \big|^2\;.
\ee

In order to minimize her probability of remaining undetected, the
eavesdropper should use a POVM and set of states that maximize
(\ref{def:Fnew}). This leads to the following definition:
\begin{defn}
Let ${\cal E} = \{p_i, | \psi_i \ket \}$ be an ensemble of states.
The {\it accessible fidelity} of $\cal E$ is defined to be
\be\label{def:acc.fid}
F({\cal E}) = \sup_{\{E_b\}} \,\, \sup_{\{|\phi_b \ket\}} \,\,
\sum_{i} \sum_{b} p_i \,\, \bra \psi_i | E_b |\psi_i \ket \, \, \big|
\bra \psi_i | \phi_b \ket \big|^2\;.
\ee
\end{defn}

Since $F({\cal E})$ is the pointwise maximum of functions that are
linear in the weights $p_i$, it is a convex function of the $p_i$.
Because the set of possible weights $\{p_i\}$ is convex (more
precisely, a simplex), the maximum value of $F({\cal E})$ over all
weights is achieved in an extreme point of the simplex \cite{rock}.
These points are characterised by one of the $p_i$ being 1 and all
the others being 0. Thus
\bee
\max_{\{p_i\}}F({\cal E})
&=&
  \max_i \sup_{\{E_b\}} \,\, \sup_{\{|\phi_b \ket\}} \,\,
  \sum_{b} p_i \,\, \bra \psi_i | E_b |\psi_i \ket \, \, \big| \bra
  \psi_i | \phi_b \ket \big|^2 \\
&=& 1\;.
\eee
The optimum is achieved by taking $\{E_b\} = \{I\}$ and
$|\phi_b\ket=|\psi_i\ket$ (for any choice of $i$).  Hence the maximum
of $F({\cal E})$ over all ensembles is not particularly interesting.

On the other hand, there are nontrivial lower bounds for the
accessible fidelity as a function of the $\{p_i\}$~\cite{fuchs}.  In
particular, the quantumness of a set of states $\{| \psi_i \ket\}$
provides an intrinsic and nontrivial character for the set itself:
\begin{defn}
The {\it quantumness} of a collection of states $\{| \psi_i \ket\}$
is defined to be
\be\label{def:qua.ness}
Q\Big(\{| \psi_i\ket\}\Big) = \inf_{\{p_i\}} \,\, F(\{p_i, | \psi_i
\ket\})\;.
\ee
\end{defn}
The quantumness specifies the best use that can be made of a set of
states for revealing the existence of an eavesdropper:  It is an
inverted measure, the smaller the quantumness, the greater the
departure from classical characteristics (since in the classical
world an unconstrained eavesdropper cannot be detected at all).

The purpose of this paper is to show that both the accessible
fidelity and the quantumness satisfy an important multiplicativity
property for product structures. To be specific, given two ensembles
${\cal E}_1 = \{p_i, |\psi_i \ket\}$ and ${\cal E}_2 = \{q_j,
|\theta_j\ket\}$, define the product ensemble ${\cal E}_1 \ot {\cal
E}_2$ by
\be\label{def:prod}
{\cal E}_1 \ot {\cal E}_2 = \{p_i q_j, |\psi_i \ket \ot |\theta_j
\ket \}\;.
\ee
We prove the following two theorems:
\begin{thm}\label{thm1}
For any ensembles ${\cal E}_1$ and ${\cal E}_2$,
\be\label{mult-F}
F({\cal E}_1 \ot {\cal E}_2) = F({\cal E}_1) \,\, F({\cal E}_2)\;.
\ee
\end{thm}
and
\begin{thm}\label{thm2}
For any collections $\{| \psi_i \ket \}$ and $\{| \theta_j \ket\}$,
\be\label{mult-Q}
Q\Big(\{| \psi_i \ket \ot | \theta_j\ket\}\Big) = Q\Big(\{| \psi_i
\ket\}\Big) \,\, Q\Big(\{| \theta_j\ket\}\Big)\;.
\ee
\end{thm}

The significance of these these theorems is the following.  In the
first case, imagine not a single shot through the eavesdropping
channel, but rather a source that repeatedly generates states from
the ensemble $\cal E$.  One could imagine a smart eavesdropper who
saves up multiple signals before performing her measurement on the
chance that it will help her remain undetected. Our first theorem
shows that this more complicated strategy provides no help. The
second theorem makes a statement about optimal uses of an alphabet.
It says, given a state preparation device that can only prepare
states from a given collection $\{| \psi_i \ket \}$, it is never in
the sender's interest to generate correlations between separate
transmissions.  In this way, quantumness is quite distinct from a
channel capacity.  For in contrast to channel capacity---where
introducing correlation is generally necessary for achieving
it---eavesdropping detection prefers uncorrelated signals. Theorem 1
and 2 together support the notion that accessible fidelity and
quantumness are intrinsic properties of an ensemble and its
underlying set of states.

The remainder of the paper is organized as follows.  In Section 2 we
lay out the basic ingredients required for proving the theorems.
Following that, in Section 3 we prove Theorem 1, and in Section 4 we
prove Theorem 2.  We conclude in Section 5 with a small discussion
about the potential implications of this work.  In an Appendix we
give a new proof of the multiplicativity of the maximal $\infty$-norm
for entanglement breaking channels.

\section{Ingredients of the proof of Theorem \ref{thm1}}
We describe here the two  principal ingredients in the proof. The
first ingredient is an application of the duality principal of convex
analysis, which allows the accessible fidelity to be rewritten as an
infimum over affine functions which majorize its value on the pure
states. Similar ideas have been exploited recently in other areas of
quantum information theory, in particular in the work on equivalence
of additivity questions~\cite{ka,Shor2}.

The second ingredient is an additivity result for a particular class
of completely positive maps known as entanglement-breaking maps. This
property was first established by Shor for the minimal entropy of
maps \cite{Shor}, and later extended to the noncommutative $p$-norms
for all $p \geq 1$ \cite{K}.

To describe the first ingredient, it is convenient to define the
following completely positive map $\Phi$ associated with an ensemble
${\cal E} = \{p_i, |\psi_i \ket\}$:
\be\label{def:Phi}
\Phi(\rho) = \sum_i p_i \Pi_i \rho \Pi_i\;,
\ee
where $\Pi_i=|\psi_i\ket\bra\psi_i|$. Then the accessible fidelity of
an ensemble (\ref{def:acc.fid}) can be rewritten in terms of $\Phi$:
\be
F({\cal E}) &=& \sup_{\{E_b\}} \,\, \sup_{\{|\phi_b \ket\}} \,\,
\sum_{i} \sum_{b} p_i \,\, \bra \psi_i | E_b |\psi_i \ket \, \,
\big| \bra \psi_i | \phi_b \ket \big|^2 \nonumber \\
&=& \sup_{\{E_b\}} \sum_b \,\, \sup_{|\phi_b \ket} \,\,
\bra\phi_b|\Phi(E_b)|\phi_b\ket \nonumber \\
&=& \sup_{\{E_b\}} \sum_{b} \, ||\Phi( E_b )||\;. \label{rewF}
\ee
Furthermore, given a POVM $\{E_b\}$, we associate to it an ensemble
of states $\{\alpha_b, \sigma_b\}$, given by
\be\label{def:a-s}
\alpha_b = {1 \over d} \, \tr (E_b)\;, \quad\quad \sigma_b = {1 \over
d \alpha_b} \, E_b\;,
\ee
where $d$ is the dimension of the state space. This defines a 1--1
correspondence between POVM's and ensembles whose average state is
${1 \over d} I$. Hence (\ref{rewF}) can be rewritten as a sup over
such ensembles, that is
\be\label{rewriteF}
F({\cal E})  = d \, \sup \Big{\{} \sum_{b} \alpha_b \,
||\Phi(\sigma_b)|| \,
  : \, \sum_b \alpha_b\, \sigma_b = {1 \over d} I \Big{\}}\;.
\ee

Introduce the following function on states:
\be\label{def:g}
g(\rho) = ||\Phi(\rho)||\;.
\ee
This function is obviously convex.
The concave closure of $g$ is defined as follows:
\be\label{def:g^}
{\hat g}(\rho) = \sup_{\{\alpha_b, \sigma_b\}} \Big{\{}\sum_{b}
\alpha_b \, g(\sigma_b) \, : \, \sum \alpha_b\, \sigma_b = \rho
\Big{\}}\;.
\ee
The concave closure of a function $g$ is the smallest concave
function on the set of all states that coincides with $g$ on the pure
states. Comparing with (\ref{rewriteF}) we can see that
\be\label{F1}
F({\cal E}) = d \, {\hat g}\Big({1 \over d} I\Big)\;.
\ee
By the dual formulation of the concave closure, ${\hat g}$ can also
be expressed as the infimum over all affine functions that dominate
$g$ \cite{rock,ka}; that is
\be\label{g^-inf}
{\hat g}(\rho) = \inf_{X} \Big{\{} \tr (X \rho) \, : \, \tr (X |\psi
\ket \bra \psi|) \geq g(|\psi \ket \bra \psi|), \, {\rm all} \, |\psi
\ket \Big{\}}\;,
\ee
where the infimum runs over all self-adjoint matrices.

Without loss of generality we can assume that the signal states span
the state space, so that any $X$ satisfying the conditions in
(\ref{g^-inf}) must be positive definite. We denote by ${\cal F}(g)$
the collection of matrices which satisfy these conditions, and call
this the {\it feasible} set for $g$. So
\be\label{g^2}
{\hat g}(\rho) = \inf_{X \in {\cal F}(g)} \tr (X \rho)\;.
\ee

For the second ingredient, recall that a completely positive map
$\Psi$ is entanglement breaking \cite{Shor} if it can be written in
the form
\be \label{eq:holv}
\Psi(\rho) = \sum_{k} \, R_k \, \tr (X_k \rho)\;,
\ee
where $\{R_k\}$ and $\{X_k\}$ are positive semidefinite.
Comparing with (\ref{def:Phi}) we can see that $\Phi$ is entanglement
breaking, where
\be
R_k = p_k \, \Pi_k\;, \quad\quad X_k = \Pi_k\;.
\ee

For any $p \geq 1$, the maximal $p$-norm of a CP map $\Omega$ is
defined by
\be
\nu_{p}(\Omega) = \sup_{|\psi\ket} || \Omega(|\psi \ket \bra \psi |)
||_p \;,
\ee
where the $p$-norm of a matrix $A$ is defined by
\be\label{def:p-norm}
||A||_p = \Big(\tr \left( A^{*} A \right)^{p/2} \Big)^{1/p}\;.
\ee
The minimal output entropy of a trace preserving CP map is equal to the
derivative of the maximal $p$-norm at $p=1$:
\be
S_{\rm min}(\Omega) = {d \over dp} \, \nu_{p}(\Omega) \,
\Big{|}_{p=1}\;.
\ee

Shor proved that the minimal output entropy of a product channel is
additive, provided that at least one of the channels is entanglement
breaking \cite{Shor}. It was later shown that the maximal $p$-norm of
such a product channel is always multiplicative, for any $p \geq 1$
\cite{K}. In fact, with a slight modification of the proof of
\cite{K} one can show that multiplicativity also holds for general CP
maps, not necessarily trace-preserving ones. In this paper we will
make use of this latter result for the case $p=\infty$. The proof
presented in \cite{K} uses the powerful Lieb-Thirring inequality
\cite{LT} to derive the result for all $p$. It turns out that for the
case $p = \infty$ there is a simpler method of proof which does not
need this level of sophistication. Therefore we state this case as a
separate Lemma below, and present its proof in the Appendix.
\begin{lem}\label{EB}
Let $\Phi$ be an entanglement-breaking CP map, and let $\Omega$ be
any other CP map. Then
\be
\nu_{\infty}(\Phi \ot \Omega) = \nu_{\infty}(\Phi) \,
\nu_{\infty}(\Omega)\;.
\ee
\end{lem}
\section{Proof of Theorem \ref{thm1}}
First we note that the inequality
\be
F({\cal E}_1 \ot {\cal E}_2) \geq F({\cal E}_1) \, F({\cal E}_2)
\ee
follows immediately from the definition (\ref{def:F}), since the
fidelity of the product ensemble ${\cal E}_1 \ot {\cal E}_2$ can only
decrease by restricting to product POVM's and product states
$\phi_b$. So the Theorem reduces to proving the inequality
\be\label{need1}
F({\cal E}_1 \ot {\cal E}_2) \leq F({\cal E}_1) \, F({\cal E}_2)\;.
\ee

Let $\Phi_1$ and $\Phi_2$ denote the CP maps defined as in
(\ref{def:Phi}) for the two ensembles ${\cal E}_1$ and ${\cal E}_2$.
It follows that the corresponding CP map for the product ensemble
${\cal E}_1 \ot {\cal E}_2$ is the product map $\Phi_1 \ot \Phi_2$.
As in (\ref{def:g}) we define the associated functions
\be
g_i(\rho) = ||{\Phi}_i(\rho)||, \quad i=1,2\;, \quad g_{12}(\rho) =
||(\Phi_1 \ot \Phi_2) \, (\rho)||\;.
\ee

Now recall (\ref{F1}).  This implies
\be
F({\cal E}_i) = d_i \, {\hat g}_i \, \Big({1 \over d_i} \, I\Big),
\quad i=1,2\;,
\ee
where $d_i$ is the dimension of the state space for the ensemble
${\cal E}_i$. From (\ref{g^2}) it follows that there are optimal
self-adjoint matrices $X_1$ and $X_2$ belonging to the feasible sets
for $g_1$ and $g_2$, respectively, such that
\be\label{X-min}
F({\cal E}_i) = \tr \, (X_i), \quad i=1,2\;,
\ee
and also that
\be\label{F_12}
F({\cal E}_1 \ot {\cal E}_2) = \inf_{X_{12} \in {\cal F}(g_{12})} \tr
\, (X_{12})\;.
\ee
Assuming that $X_1 \ot X_2 \in {\cal F}(g_{12})$, it follows that
\be
F({\cal E}_1 \ot {\cal E}_2) \leq \tr \, (X_1 \ot X_2) = \tr(X_1) \, \tr(X_2)
\ee
which gives the desired inequality (\ref{need1}).

So we are left with proving the assumption:
\begin{lem}\label{feas}
Let $X_1$ and $X_2$ belong to the feasible sets of $g_1$ and $g_2$
respectively. Then $X_1 \ot X_2$ belongs to the feasible set of
$g_{12}$.
\end{lem}
\noindent{\it Proof}:
Recall that every matrix in the feasible set of $\cal E$ is positive
definite. Given the two  matrices $X_i\in {\cal F}(g_{i})$, $i=1,2$,
define the entanglement-breaking CP maps $\Omega_{1}$ and
$\Omega_{2}$ by
\be
\Omega_{i} (\rho) = {\Phi}_{i} \,\Big( X_{i}^{-1/2} \rho X_{i}^{-1/2}
\Big), \quad i=1,2\;.
\ee
The feasibility of $X_i$ means that for all pure states $|\psi\ket$:
\be
\tr[ X_i | \psi \ket \bra \psi |] \ge g_i(| \psi \ket \bra \psi |) =
||\Phi_i(| \psi \ket \bra \psi |)||\;.
\ee
Substituting
\be
|\psi\ket = X_i^{-1/2} |\phi\ket\;,
\ee
it follows that for any pure state $| \phi \ket$
\be
|| \Omega_{i} (| \phi \ket \bra \phi |) || \leq 1, \quad i=1,2
\ee
and hence that
\be
{\nu}_{\infty}(\Omega_{i}) \leq 1\;.
\ee
Hence from Lemma \ref{EB} we get
\be
|| (\Omega_{1} \ot \Omega_{2}) (| {\psi}_{12} \ket \bra {\psi}_{12}
|) || \leq 1=\tr\; | {\psi}_{12} \ket \bra {\psi}_{12}|
\ee
for any pure state $| {\psi}_{12} \ket$. This implies in turn that
\be
|| (\Phi_{1} \ot \Phi_{2}) (\rho_{12}) || \leq
\tr [ (X_1 \ot X_2)\rho_{12}]
\ee
for any bipartite state $\rho_{12}$. Hence $X_1 \ot X_2$ is in the feasible set
for $g_{12}$.
\qed
\section{Proof of Theorem \ref{thm2}}
First, by restricting to
product distributions it follows immediately from Theorem \ref{thm1} that
\be\label{leq}
Q\Big(\{| \psi_i \ket \ot | \theta_j \ket\}\Big) \leq Q\Big(\{|\psi_i
\ket\}\Big) \,\, Q\Big(\{| \theta_j \ket\}\Big)\;.
\ee
So it sufficient to prove the bound in the other direction.

We need to prove that for any joint distribution $\{p_{ij}\}$ on the
collection of product states $\{|\psi_i \ket \ot | \theta_j \ket\}$,
we have
\be\label{suff.bound}
F(\{p_{ij}, | \psi_i \ket \ot | \theta_j \ket\}) \geq Q\Big(\{|\psi_i
\ket\}\Big) \,\, Q\Big(\{| \theta_j \ket\}\Big)\;.
\ee
Indeed, taking the infimum over all distributions $\{p_{ij}\}$ in
(\ref{suff.bound}) gives the inequality
\be\label{geq}
Q\Big(\{| \psi_i \ket \ot |\theta_j \ket\}\Big) \geq Q\Big(\{| \psi_i
\ket\}\Big) \,\,Q\Big(\{| \theta_j \ket\}\Big)\;,
\ee
which together with (\ref{leq}) yields (\ref{mult-Q}).

Since the accessible fidelity is a supremum over POVMs,
to prove (\ref{suff.bound}), it is enough to find a particular
POVM $\{M_{b,c}\}$ such that
\be\label{suff2}
\sum_{b,c} \, \Big|\Big| \sum_{i,j} p_{ij} \, \Big(\Pi_{i} \ot
{\tilde \Pi}_j \Big) \, M_{b,c} \, \Big(\Pi_{i} \ot {\tilde
\Pi}_j\Big) \Big| \Big| \geq Q\Big(\{| \psi_i \ket\}\Big) \,\,
Q\Big(\{|\theta_j \ket\}\Big)\;,
\ee
with $\Pi_i = | \psi_i \ket \bra \psi_i|$ and ${\tilde \Pi}_j = |
\theta_j \ket \bra \theta_j |$. It will become clear further on why
we have equipped $M_{b,c}$ with two indices.

The POVM $\{M_{b,c}\}$ is constructed in two steps.
First, define the marginal distribution
\be
p_i = \sum_{j} p_{ij}.
\ee
Let $\{E_b\}$ be an optimal POVM realising $F(\{p_i,|\psi_i\ket\})$,
so that
\be\label{def:E}
F(\{p_i, |\psi_i \ket\}) = \sum_b \, \Big|\Big| \sum_{i} p_{i} \,
\Pi_{i} \,E_b \, \Pi_{i} \Big| \Big|\;.
\ee
For each $b$, let $|\phi_b \ket$ be the dominating eigenvector of
$\sum_{i} p_{i} \, \Pi_{i} \,E_b \, \Pi_{i}$
so that
\be
\bra \phi_b | \sum_i p_{i} \, \Pi_{i}\, E_b \, \Pi_{i} | \phi_b \ket
= \Big|\Big| \sum_{i} p_{i} \,\Pi_{i}  \, E_b \, \Pi_{i} \Big|
\Big|\;.
\ee

Define for every $b$ a new distribution $\{q_{b,j}\}_j$ by
\be\label{def:q-j}
q_{b,j} = \frac{1}{N_b} \,\, \sum_{i} p_{ij} \, \bra \phi_b 
|  \Pi_{i}  \,E_b \, \Pi_{i} | \phi_b \ket,
\ee
where the normalisation constant is
$N_b=\sum_j \sum_i p_{ij} \, \bra \phi_b |  \Pi_{i}  \,E_b \, \Pi_{i} | 
\phi_b \ket$.
Remark that with this notation,
\be
F(\{p_i, |\psi_i \ket\}) = \sum_b \, N_b.
\ee

For each $b$ let $\{F_{b,c}\}$ be an optimal POVM realising
$F(\{q_{b,j},|\theta_j\ket\})$, so that
\be\label{def:Fter}
F(\{q_{b,j}, |\theta_{j}\ket \})
= \sum_{c} \, \Big|\Big| \sum_{j} q_{b,j} \,{\tilde \Pi}_{j} \, F_{b,c} \, 
{\tilde \Pi}_{j} \Big| \Big|,
\ee
and, for each $b,c$, let $| \chi_{b,c} \ket$ be the dominating eigenvector of
$\sum_{j} q_{b,j} \, {\tilde \Pi}_{j} \, F_{b,c} \, {\tilde \Pi}_{j}$,
so that
\be
\bra \chi_{b,c} | \sum_{j} q_{b,j} \, {\tilde \Pi}_{j} \, F_{b,c} \, 
{\tilde \Pi}_{j} | \chi_{b,c} \ket
=      \Big|\Big| \sum_{j} q_{b,j} \, {\tilde \Pi}_{j} \, F_{b,c} \, 
{\tilde \Pi}_{j} \Big| \Big|.
\ee
Hence,
\be
\lefteqn{F(\{q_{b,j}, |\theta_{j}\ket \})} \nonumber\\
&=&
\sum_c \, \bra \chi_{b,c} | \sum_{j} q_{b,j} \, {\tilde \Pi}_{j} \, F_{b,c} 
\, {\tilde \Pi}_{j} | \chi_{b,c} \ket
\nonumber\\
&=& \sum_c \frac{1}{N_b}\sum_{i,j} p_{ij} \,
\bra \phi_b | p_{i} \, \Pi_{i}\, E_b \, \Pi_{i} | \phi_b \ket \,\bra
\chi_{b,c} | q_{b,j} \, {\tilde \Pi}_{j} \, F_{b,c} \, {\tilde
\Pi}_{j} | \chi_{b,c} \ket
\nonumber\\
&=& \frac{1}{N_b}\sum_c \bra \phi_b\ot\chi_{b,c}| \sum_{ij} p_{ij}
(\Pi_i\ot{\tilde\Pi}_j) \, E_b\ot F_{b,c}\, (\Pi_i\ot{\tilde\Pi}_j)
|\phi_b\ot\chi_{b,c}\ket \nonumber \\
&\le& \frac{1}{N_b}\sum_c \Big|\Big|
\sum_{ij} p_{ij} (\Pi_i\ot{\tilde\Pi}_j) \, E_b\ot F_{b,c}\,
(\Pi_i\ot{\tilde\Pi}_j) \Big|\Big|\;.
\ee

Define yet another distribution $\{r_b\}$ by $r_b=N_b/\sum_a N_a$, then
\be
\lefteqn{\sum_b r_b F(\{q_{b,j}, |\theta_{j}\ket \})} \nonumber\\
&\le& \frac{1}{\sum_a N_a}\sum_{b,c}
\Big|\Big| \sum_{ij} p_{ij} (\Pi_i\ot{\tilde\Pi}_j) \, E_b\ot
F_{b,c}\, (\Pi_i\ot{\tilde\Pi}_j) \Big|\Big|\;.
\ee

The POVM $\{M_{b,c}\}$ is now defined by
\be\label{def:M}
M_{b,c} = E_b \ot F_{b,c}\;.
\ee
 From combining all the above it follows that
\be
\lefteqn{F(\{p_i, | \psi_i \ket\}) \,
\sum_b r_b F(\{q_{b,j}, |\theta_{j}\ket \})} \nonumber \\
& \leq & \sum_{b,c} \, \Big|\Big|
\sum_{ij} p_{ij} (\Pi_i \ot{\tilde \Pi}_{j}) M_{b,c}
(\Pi_i \ot {\tilde \Pi}_{j}) \Big|\Big|\;. \label{suff3}
\ee
Now the definition of quantumness implies that
\be
F(\{p_i, |\psi_i \ket\}) \geq Q(\{| \psi_i \ket\})
\ee
and
\be
F(\{q_{b,j}, | \theta_{j} \}) \geq Q(\{| \theta_{j}\ket \})\;,
\ee
so that
\be
\sum_b r_b F(\{q_{b,j}, | \theta_{j} \}) \geq Q(\{| \theta_{j}\ket
\})\;,
\ee
and together with (\ref{suff3}) this gives (\ref{suff2}).
\qed
\section{Discussion}

The present work clarifies two things.  First, that both accessible
fidelity and quantumness should be written in ``single-letterized''
expressions, as they were originally proposed. Second, Theorem 1 {\it
may\/} lend some evidence to the idea that collective eavesdropping
strategies need not be considered in a full quantum eavesdropping
analysis after all---an idea that has been toyed with in the
past~\cite{Gisin01}. If true, this would significantly relieve the
technological requirements for operational systems in which
unconditional security is sought.

Beyond this, one of the authors (CAF) is hopeful that these
measures---particularly quantumness---will be useful to a certain
line of attack in quantum foundations~\cite{Fuchs02}.  In that
approach, a quantum state represents not an intrinsic property of a
system, but rather an observer's information---namely, the best
information that can be had given that the components of the world
have a certain fundamental sensitivity to the touch.

\begin{acknowledgement}
KA and AW thank CNRI in Dublin for its hospitality, where part of
this work was performed. CF and CK were supported in part by Science
Foundation Ireland under the National Development Plan.  CK was also
supported in part by National Science Foundation Grant DMS-0101205.
AW was supported by the U.K. Engineering and Physical Sciences
Research Council.

\end{acknowledgement}
\section{Appendix: Proof of Lemma \ref{EB}}
The proof follows the line of argument presented in \cite{K},
but replacing the Lieb-Thirring inequality with a simpler bound
for the operator norm.

We show here that entanglement-breaking CP maps satisfy
multiplicativity of the maximal $\infty$-norm. The maximal
$\infty$-norm of a CP map $\Omega$ is defined as
\be\label{def:nu}
\nu_{\infty}(\Omega) = \sup_{\rho} \, || \Omega(\rho) ||\;,
\ee
where the $\sup$ runs over all density matrices in the domain of $\Omega$.
It is trivial to show that
\bee
\nu_{\infty}(\Psi \ot \Omega) \geq \nu_{\infty}(\Psi) \,
\nu_{\infty}(\Omega)\;.
\eee
Simply let $\rho_1$ and $\rho_2$ be states that achieve
$\nu_{\infty}(\Psi)$ and $\nu_{\infty}(\Omega)$, respectively. Then
$\rho_1\ot\rho_2$ is not necessarily optimal for
$\nu_\infty(\Psi\ot\Omega)$, so that
\bee
\nu_{\infty}(\Psi \ot \Omega)
&\ge& || (\Psi\ot\Omega)(\rho_1\ot\rho_2) || \\
&=& || \Psi(\rho_1) ||\,\,||\Omega(\rho_2) || \\
&=& \nu_{\infty}(\Psi) \, \nu_{\infty}(\Omega)\;.
\eee
Therefore, to prove the Lemma, we only need to show that
\bee
\nu_{\infty}(\Psi \ot \Omega) \leq \nu_{\infty}(\Psi) \,
\nu_{\infty}(\Omega)\;.
\eee

To set up the notation, consider the action of the map
(\ref{eq:holv}) on a bipartite state $\rho_{12}$:
\be\label{bip}
(\Psi \ot I) (\rho_{12}) = \sum_{k=1}^K \, R_k \, \ot \, {\tr}_{1}\,
[  (X_k \ot I) \rho_{12} ]
\ee
and let
\be
\rho_{12} = (I \ot \Omega)(\tau_{12})\;.
\ee
Then
\be\label{tau}
(\Psi \ot I) (\rho_{12}) = (\Psi \ot \Omega) (\tau_{12})\;.
\ee
Define
\be\label{def:x,G}
x_{k} & = &  \tr \, [  (X_k \ot I) \tau_{12} ] \\
G'_{k} &= & {\tr}_{1} \, [  (X_k \ot I) \tau_{12} ]/x_{k} \nonumber \\
G_{k} & = & \Omega(G'_k) = {\tr}_{1} \, [  (X_k \ot I) \rho_{12}
]/x_{k}\;.
\nonumber
\ee
Then (\ref{bip}) reads
\be
(\Psi \ot I) (\rho_{12}) &=& \sum_{k=1}^K \, x_{k} R_k  \,\ot\, 
G_k  \label{bip2} \\
(\Psi \ot I) (\tau_{12}) &=& \sum_{k=1}^K \, x_{k} R_k  \,\ot\,
G'_k\;,
\label{bip2a}
\ee
where now $\{ R_k, G_k \}$ are all positive matrices, $G'_k$ is a density 
matrix
and $x_k \geq 0$.
Writing $\tau_1 = {\tr}_{2} (\tau_{12})$ for the reduced density matrix
it follows from (\ref{bip2a}) that
\be\label{Phi-red}
\Psi(\tau_1) =  \sum_{k=1}^K \, x_{k} R_k\;.
\ee
Noting that for any Hermitian matrix $X$, $X\le ||X||\,I$, we have
\be
(\Psi\ot\Omega)(\tau_{12}) &=& (\Psi \ot I) (\rho_{12}) \nonumber \\
&=&   \sum_{k=1}^K \, x_{k} R_k  \,\ot\, G_k  \nonumber \\
&\le& \sum_{k=1}^K \, x_{k} R_k \,\ot\, ||G_k||\,I \nonumber \\
&\le& (\max_k ||G_k||)\, \sum_{k=1}^K \, x_{k} R_k \,\ot\, I \nonumber \\
&=& (\max_k ||G_k||) \, \Psi(\tau_1) \,\ot\, I\;. \label{tt}
\ee
Now recollect that $G_k=\Omega(G'_k)$ and that $G'_k$ is a density matrix.
Therefore (\ref{def:nu}) implies that
\bee
||G_{k}|| \leq \nu_{\infty}(\Omega)
\eee
for any $k$. Together with (\ref{tt}) and the fact that tensoring in
the identity does not change the operator norm, this implies
\be
||(\Psi \ot \Omega) (\tau_{12})|| \leq  \nu_{\infty}(\Omega) \,
||\Psi(\tau_{1})||\;.
\ee
Using again (\ref{def:nu}) we get
\be
||(\Psi \ot \Omega) (\tau_{12}) || \le \nu_{\infty}(\Omega) \,
\nu_\infty(\Psi).
\ee
Since this bound holds for all $\tau_{12}$ it follows that
\be
\nu_{\infty}(\Psi \ot \Omega) \leq \nu_{\infty}(\Psi) \,
\nu_{\infty}(\Omega)\;.
\ee
\qed

{~~}
\end{document}